\documentclass[12pt]{article}
\usepackage{graphicx}
\usepackage{epsf}

\def\a{\alpha}
\def\b{\beta}
\def\c{\chi}

\def\eps{\varepsilon}
\def\f{\frac}
\def\g{\gamma}
\def\G{\Gamma}

\def\l{\left}
\def\m{\mu}

\def\p{\partial}

\def\r{\right}
\def\s{\sigma}
\def\t{\theta}
\def\vp{\varphi}

\def\be{\begin{equation}}
\def\ee{\end{equation}}
\def\bea{\begin{eqnarray}}
\def\eea{\end{eqnarray}}
\def\ba{\begin{array}}
\def\ea{\end{array}}

\def\mc{\mathcal}

\begin{document}
\begin{titlepage}
\begin{center}
{\Large \bf William I. Fine Theoretical Physics Institute \\
University of Minnesota \\}
\end{center}
\vspace{0.15in}
\begin{flushright}
FTPI-MINN-10/10 \\
UMN-TH-2901/10 \\
April 2010 \\
\end{flushright}
\vspace{0.15in}
\begin{center}
{\Large \bf Semiclassical calculation of an induced decay of false vacuum \\}
\vspace{0.15in}
{\bf A. Monin \\}
School of Physics and Astronomy, University of Minnesota, \\ Minneapolis, MN
55455, USA, \\
and \\
{\bf M.B. Voloshin  \\ }
William I. Fine Theoretical Physics Institute, University of
Minnesota,\\ Minneapolis, MN 55455, USA \\
and \\
Institute of Theoretical and Experimental Physics, Moscow, 117218, Russia
\\[0.1in]
\end{center}

\begin{abstract}
We consider a model where a scalar field develops a metastable vacuum state and weakly interacts with another scalar field. In this situation we find the probability of decay of the false vacuum stimulated by the presence and collisions of particles of the second field. The discussed calculation is an illustration of the recently suggested thermal approach to treatment of induced semiclassical processes.  
\end{abstract}

\end{titlepage}
\newpage

\section{Introduction}
The class of problems involving spontaneous and induced semiclassical processes with quantum fields invariably presents an interesting case study in nonperturbative calculations. The most discussed examples of such processes are the Schwinger  $e ^ + e ^ -$ pair creation in an electric field~\cite{Sauter,Heisenberg:1935qt,Schwinger},  the decay of false vacuum~\cite{vko,Coleman,Voloshin:1985id} and the breakup of metastable strings and walls~\cite{mvsw}. A semiclassical treatment of such processes can be formulated in terms of a configuration in the Euclidean space time, which is a solution to classical field equations and is called the bounce~\cite{Coleman,cc,Affleck,Dunne:2006st}. The considered processes can either proceed spontaneously or be induced by the presence of particles and by their collisions, such as in the false vacuum decay~\cite{sv,mv3,Gorsky:2005yq}, in the decay of metastable strings and walls~\cite{mvsw,mvsp}, and in the photon-stimulated Schwinger process~\cite{sgd,Dunne:2009gi}. In terms of a Euclidean-space treatment the case of induced processes requires one to consider the interaction of the particles with the semiclassical configuration. It has been recently suggested that such a consideration can be done using an indirect thermal method~\cite{mvsw,mv2009,mv2010}, which appears to be simpler than a direct calculation of a relevant correlator~\cite{mv2010} in the semiclassical background, especially if more than one quantum particle is involved.

The thermal method allows one to find the rate of the induced process by appropriately interpreting the result for the spontaneous process at a small but nonzero temperature, where the thermal effects in the rate are contributed by all the $n$-particle states rather than by the vacuum state alone.   The contribution of the states with a fixed number of particles can be extracted from the thermal rate and thus the decay rate induced by $n$ particles can be calculated. An alternative, and more direct, calculation employing correlators of field operators in the bounce background is readily tractable in the case of one particle, but becomes more cumbersome for processes with a larger number of particles. It has been demonstrated~\cite{mv2010} that both methods yield the same result for the one photon induced Schwinger pair creation.

In this paper we consider an application of both approaches to the case of interacting scalar particles in two dimensions.  We consider two interacting scalar fields in the following setup. The potential for one of the fields $\vp$ is chosen as having the shape shown in Fig.1. It is well known that the system with such potential admits quantum tunneling from the metastable to the true vacuum, therefore one can discuss the spontaneous decay of the false vacuum \cite{vko,Coleman,cc,Voloshin:1985id}. Choosing the parameter of the potential asymmetry to be small and the mass of the corresponding field $\vp$ to be much larger than that of the second field $\c$, $M \gg m$, greatly simplifies the treatment. Indeed, for a small difference of the vacuum energy density $\eps$ one can employ the so called thin wall approximation~\cite{vko,Coleman}. In this case the tunneling configuration is given by a disk of the radius 
\be
R = \f {\m} {\eps},
\label{b_rad}
\ee
with the false vacuum outside of the disk and the true vacuum inside. The extent of the transition region between the two vacua is assumed to be much shorter than the radius $R$ (the thin wall approximation), which is ensured by the condition $M \, R \gg 1$. The quantity $\m$ in Eq.(\ref{b_rad}) is the mass associated with the transition region, $\m \gg M$.
The smallness of the mass $m$ of the second field $\c$ compared to $M$ guarantees that in the leading order there is no contribution to the interaction of the bubble (bounce) with other objects due to the self-action of the field $\vp$ but only due to the exchange of the light particle $\c$.

The problem that is addressed here is that of the false vacuum decay stimulated either by the presence of one particle of the field $\c$ or by a collision between two such particles. Previously a similar problem was considered~\cite{sv,vs,Gorsky:2005yq} for the particles of the field $\vp$, which is always strongly coupled to the bounce in the sense that there are always zero and soft modes for this field localized on the bounce. Our present problem is different in that we consider the catalysis of the process by particles of a second field $\c$ that is only weakly coupled to the master field $\vp$ so that no anomalously soft modes for the field $\c$ arise on the bounce. As a result we find closed formulas for the probability of an induced vacuum decay, which describe the onset of the expected semiclassical behavior similar to that the catalysis by the particles of the master field~\cite{sv,vs,Gorsky:2005yq}, where the energy of the initial particle(s) is transfered to the tunneling degree of freedom.

Since the main purpose of the present paper is an illustration of an application of two approaches to treatment of an induced semiclassical process, we limit ourselves, for simplicity, to a model in (1+1) space-time dimensions, while a generalization to higher dimensional models is quite straightforward.

The paper is organized as follows. In Sec.2 we briefly describe the model used and reformulate the problem in the thin wall limit. In Sec.3 we find the rate of the metastable vacuum decay at small but finite temperature. Using the correspondence between the $n$-particle contribution and the rate at finite temperature we find the width of the $\c$ particle and the probability of the collision-induced false vacuum decay in Sec.4. In Sec.5 we discuss a calculation of the relevant field correlator in the bounce background, and thereby calculate anew the decay rate of a single $\c$ particle. In Sec.6 we discuss and summarize the results.

\section{Vacuum tunneling in Euclidean space time}
We consider a model describing two real scalar fields in $1+1$ dimensions with the Lagrangian  
\be
\mc{L} = \f {1} {2}\l ( \p _ \m \vp \r ) ^ 2 + \f {1} {2} \l ( \p _ \m \chi \r ) ^ 2 - \f{1} {2} 
m ^ 2 \c ^ 2 - V (\vp) - V _ {\rm int} (\vp,\c),
\label{s_act}
\ee
where the potential $V (\phi)$ has asymmetric form shown in Fig.1. For further uses this potential can be chosen as
\be
V (\vp) = \f {1} {4} \lambda \l ( \vp ^ 2 -v ^ 2 \r ) ^ 2 - \f {\eps} {2v} \l ( \vp + v \r ).
\ee
We also choose the interaction potential between the two scalar fields in such a way that it is not equal to zero only where 
the field $\vp$ differs from its vacuum values, for example
\be
V _ {\rm int} =  \a \l (  \vp ^ 2 - v ^ 2 \r ) \c = \a \rho (\vp) \c \, ,
\label{int_pot}
\ee
with $\a$ being a constant. As argued previously, the parameters of the Lagrangian are assumed to be such  that the mass of the field $\vp$ is much greater then that of the field $\c$, namely
\be
m \ll M = v \sqrt {2 \lambda} \, .
\ee
\begin{figure}[ht]
  \begin{center}
    \leavevmode
    \epsfxsize=6cm
    \epsfbox{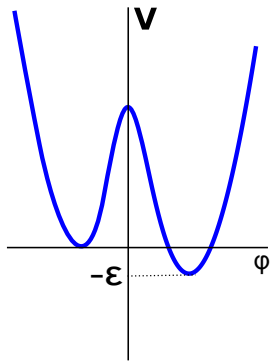}
    \caption{The asymmetric potential for the master field $\vp$.}
  \end{center}
\end{figure}
In the Euclidean time the solution for the problem of tunneling (metastable vacuum decay) can be formulated in terms of a semiclassical solution, a bounce. This is a solution to the Euclidean equations of motion interpolating between the two vacua separated by a domain wall. The decay rate is found by calculating the partition function around the bounce configuration
\be
\f {\G} {L} = \f {2} {L T} \mathrm{Im} \int \mathcal{D} \vp \mathcal {D} \c e ^ {- S [\vp,\c]},
\label{path_int_f}
\ee
where $T$, $L$ are the sizes of the system along the Euclidean time and spatial direction respectively. In the thin wall limit, when the size of the bounce is much greater than the thickness of the wall (the inverse mass of the field $\vp$)  $R = \m / \eps \gg 1/M$, one can consider only the soft modes of the wall. Then its dynamics is described in purely geometrical terms, namely the shape of the wall, which can be represented in polar coordinates by the function $r(\t)$ (closed trajectory corresponding to the distorted bounce configuration). Therefore the Euclidean action of the system can be written in the following form
\be
S = \m \ell - \eps \mathcal {A} + \int d ^ 2 x \l[ \f {1} {2} \l ( \p _ \m \chi \r ) ^ 2 + \f{1} {2} 
m ^ 2 \c ^ 2 + \rho ^{(r)} \c \r ],
\label{e_act}
\ee
where $\ell = \int d \t \sqrt {\dot {r} ^ 2 + r ^ 2}$, $\mathcal {A} = \f {1} {2} \int d \t r ^ 2$ are the length of the bounce boundary and the area it encircles, and the dot stands for the derivative with respect to $\t$. The density $\rho ^ {(r)}$ is proportional to the coefficient $\a$ and it plays the role of the source for the $\c$ field. It is obvious that in the discussed approximation the area, where the value of field $\vp$ substantially differs from the vacuum, is very thin and the support of the function $\rho ^ {(r)}$ coincides with the shape of the boundary of the bounce. Hence, it has the form of a surface $\delta$-function
\be
\rho ^ {(r)} (t,x) = g \, \delta (\sqrt{t ^ 2 + x ^ 2} - r (\t)), ~~~~ g \propto \a.
\label{dens}
\ee
As a result the partition function can be expressed as an integral over  closed trajectories describing the shape of the boundary
\be
\f {\G} {L} = \f {2} {L T} \mathrm{Im} \int \mathcal{D} r \mathcal {D} \c e ^ {- S [r,\c]},
\label{path_int}
\ee
with the action $S$ derived from Eq.(\ref{e_act}). At a small constant $g$ the latter expression can be expanded in a power series in $g$. Therefore at a small coupling between the scalar fields one can find the decay rate of the false vacuum in terms of an expansion in powers of $g$. The decay induced by $n$ particles of the field $\c$ arises in the order $g^{2n}$, while the spontaneous vacuum transition proceeds at $g \to 0$. The rate of the spontaneous decay is found from calculating the action on the bounce configuration (\ref{b_rad}) and the determinant corresponding to the integration over small fluctuations around the tunneling trajectory, and is given by~\cite{Voloshin:1985id}
\be
\f {\G} {L} = \f {\eps} {2 \pi} \exp \l [ - \f {\pi \m ^ 2} {\eps} \r ].
\label{zero_t}
\ee
In the following sections we calculate the rate of the induced decay by using two approaches, discussed in the Introduction.

\section{False vacuum decay in thermal bath}
As is already mentioned, one can use a thermal approach to extract~\cite{mvsw,mvsp,mv2009} from the expression for the total process rate at a small but finite temperature the contribution of the processes induced by a specific number of particles present in the initial state. Therefore we start with calculating the total rate of the false vacuum decay in a thermal bath. It is quite clear that at a sufficiently low temperature only the contribution of states containing particles of the field $\c$ are relevant, while the contribution from the states with the $\vp$ excitations are suppressed by the Gibbs factor $e ^ {-M/T}$.

At a low temperature one can consider the same effective Euclidean action (\ref{e_act}), except that now the system lives on the cylinder with the period equal to the inverse temperature $\b = 1/T$. In other words, one may consider the system living on the $(t,x)$ plane with periodic boundary conditions for the Euclidean time coordinate $t \to t+\b$ (see Fig.2).
\begin{figure}[ht]
  \begin{center}
    \leavevmode
    \epsfxsize=10cm
    \epsfbox{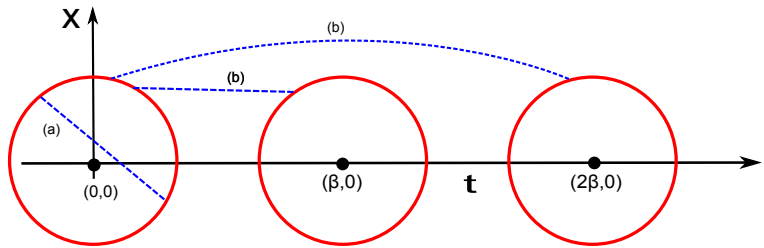}
    \caption{Periodic copies of the bounce and their interactions through the field $\c$. The self interaction (a) within one copy does not depend on the period $\b$, while the interaction (b) between different copies does depend on $\b$ and describes the thermal effects. }
  \end{center}
\end{figure}
It is worth noting that the use of the action from Eq.(\ref{e_act}) is justified as long as the temperature satisfies the condition $T \ll M$. However, we consider still much lower temperatures, such that a stronger condition $T \ll \eps/\m$ is satisfied, which drastically simplifies the calculations. At such a small temperature (equivalently, at a large period $\b$) one still may use the same bounce configuration to calculate the partition function (\ref{path_int}). The zero temperature result corresponds to the leading order contribution at $\b \to \infty$ and is given by Eq.(\ref{zero_t}). The next to leading order corrections arise in fact from the exchange of the $\c$ particle, since at $T < 2 \, R$ there is no bounce deformation~\cite{Garriga:1994ut}. The effect of the exchange of $\c$ can be classified into two types. One is the self action of the bounce independent of the temperature, it is labeled by (a) in the Fig.2. The second, labeled as (b), is the interaction of the two copies of the bounce separated along the Euclidean time by $\b$, $2\b$, and so on. In fact, it is still a part of the self interaction of one and the same bounce, living on a cylinder, and this part clearly depends on the period of the cylinder and disappears when $\b$ goes to infinity. We restrict our interest to the second type of contribution, since this is the only one that produces a nontrivial temperature dependence.  

The contribution to the action due to the interaction of two copies of the bounce separated by $\b$ can be found as usual by using the expression for the propagator of the scalar particle with mass $m$. In $1+1$ dimensions the propagator $D$ satisfies the equation
\be
-\Delta D ( t, x ) + m ^ 2 D (t,x) = \delta ^ { ( 2 ) } (t,x),
\ee
and is given by
\be
D(t,x) = \f {1} {2 \pi} K _ 0 (mr),
\ee
where $K _ 0 (mr)$ is the modified Bessel function of the second kind, and $r = \sqrt {t ^ 2 + x ^ 2} $.  The expression for the part of the action due to the interaction between two copies of the  bounce then takes the form
\be
\Delta S _ {(1)} = - \f {1} {2} \int \rho ^ B (t _ 1, x _ 1) D (t _ 2 - t _ 1,x _ 2 - x _ 1) 
\rho ^ B (t_2-\b,x_2) d t _ 1 d x _1 \, d t _ 2 d x _2,
\label{g_act_cor}
\ee
where $\rho ^ B$ is the density (\ref{dens}) for the bounce located at the origin
\be
\rho ^ B (t,x) = g \, \delta (\sqrt{t ^ 2 + x ^ 2} - R).
\label{b_dens}
\ee
The shift in the argument of the second function $\rho ^ B (t_ 2 - \b, x _ 2)$ clearly corresponds to the second copy of the bounce placed at $(\b,0)$. The extra factor of one half in the expression (\ref{g_act_cor}) is due to the apparent double counting: the expression without this factor corresponds to the interaction between the two copies of the circle, while we need to consider the change in the action per one bounce (per one period).

In order to find the integral in Eq.(\ref{g_act_cor}) one can readily notice that the function
\be
\c(t _ 2 , x_ 2) = \int  D _ 0 (t _ 2 - t _ 1, x _ 2 - x _ 1) \rho ^ B (t _ 1, x _ 1) d t _ 1 d x _ 1
\ee
satisfies the equation
\be
\l ( -\Delta + m ^ 2 \r ) \c (t,x) = g \delta (\sqrt {t ^ 2 + x ^ 2 } - R).
\label{G_r}
\ee
Let $(r _ i, \t _ i)$ be the polar coordinates corresponding to the  Cartesian ones $(t _ i, x _ i)$. The solution to the equation (\ref{G_r}) can then be found as
\be
\c(r) = \l \{
\ba{ccc}
g R \, I _ 0 (m R) \, K _ 0 (mr),~~~~ r > R, \\
g R \, I _ 0 (mr) \, K _ 0 (mR),~~~~ r < R, 
\ea
\r.
\ee
and the integration over the first circle then reduces to the following
\be
\int K _ 0 \l ( m \sqrt {r _ 2 ^ 2 + R ^ 2 - 2 r _ 2 R \cos \t _ 1} \r ) d \t _ 1 = 
2 \pi \, I _ 0 (m R) \, K _ 0 (m r _ 2)\, , 
\ee
where $I _ 0 (m R)$ is the modified Bessel function of the first kind.
Performing the further integration over $r _ 2$ and $\t _ 2$ in Eq.(\ref{g_act_cor}) and summing over the contributions from the interaction with all copies  one finally finds
\be
\f {\G_ T} {L} = \f {\eps} {2 \pi} \exp \l [ - \f {\pi \m ^ 2} {\eps} - \Delta S \r ],
\label{nzero_t}
\ee
with the total temperature dependent part of the action given by the following expression
\be
\Delta S = - 2 \pi g ^ 2 R ^ 2 I _ 0 ^ 2 (m R) \sum _ {n = 1} ^ {\infty} K _ 0 (m \b n),
\label{D_act}
\ee
Expanding the expression (\ref{nzero_t}) in $\Delta S$ at large $\b$ one can find the low temperature behavior of the thermal corrections to the decay rate: 
\be
\f {\G_ T} {L} = \f {\eps} {2 \pi} \exp \l [ - \f {\pi \m ^ 2} {\eps} \r ] \l ( 1 + 2 \pi g ^ 2 R ^ 2 I _ 0 ^ 2 (m R) \sqrt { \f {\pi} {2 m \b} } \, e ^ {-m \b} + \dots \r ).
\ee
As expected, these corrections are exponentially suppressed in the parameter $m/T$.

\section{Induced false vacuum decay}
The enhancement of the tunneling at a finite temperature arises through the stimulation of the process by the particles present in the bath.
The dependence of the rate of the process induced by the $\c$ particles on their energies then translates into the dependence on the temperature T after averaging over the thermal distribution of the particles with the standard density function
\be
n ({\vec k}) = \f {1} {e ^ {E _ k \b} - 1}\, , 
\ee
where the energy is given by $E _ k = \sqrt { {\vec k } ^ 2 + m ^ 2}$.
The number of particles involved in each of the microscopic processes can
be readily identified by the power of the factor $g ^ 2$. Since the thermal correction (\ref{D_act}) in the
action is proportional to $g ^ 2$, the $n$-particle contribution to the thermal rate is given by the
$n$-th power of $\Delta S$ in the expansion of the factor $\exp \l ( - \Delta S\r )$ in the expression (\ref{nzero_t}). In particular, the one and two-particle contributions to the decay rate in a thermal state are given by
\bea
\f {\G_ {1 \c}} {L} & = & \f {\eps} {2 \pi} \exp \l [ - \f {\pi \m ^ 2} {\eps} \r ] 
\l ( -  \Delta S   \r ), \nonumber \\
\f {\G_ {2 \c}} {L} & = & \f {\eps} {2 \pi} \exp \l [ - \f {\pi \m ^ 2} {\eps} \r ] 
\f {1} {2} \l (   \Delta S  \r ) ^ 2.
\label{1_2_contr}
\eea
On the other hand, the one-particle contribution can be found in terms of the width $\g _ \c$ of the particle $\c$ associated with the destruction of the false vacuum by the particle's presence:
\be
\f {\G_ {1 \c}} {L} = \int _{-\infty}^{+\infty}  \f {d k} {2 \pi} \f {\g _ \c \, m} {\sqrt { { k } ^ 2 + m ^ 2} } \, n({k}) = 
\f {m \g _ \c} { \pi} \sum _ {n = 1} ^ \infty \int _ {m \b} ^ \infty \f {d x} {\sqrt{x ^ 2 - (m \b) ^ 2}} \, e ^ {-n x} = \f {m \g _ \c} { \pi} \sum _ {n = 1} ^ \infty K _ 0 (m \b n).
\label{G1_t}
\ee
The origin of the Lorentz factor $m/\sqrt{{\vec k} ^ 2 + m ^ 2}$ is rather straightforward: the width $\g _ \c$ is measured in its rest frame, while we consider the process in the laboratory frame. Comparing the two results (\ref{1_2_contr}) and (\ref{G1_t}) we find the width
\be
\g _ \c = \f { \pi \, \eps} {m} \, g ^ 2 R ^ 2 I _ 0 ^ 2 (m R) \, \exp \l [ - \f {\pi \m ^ 2} {\eps} \r ] \, .
\label{t_width}
\ee

Similarly, one can express the two-particle contribution from (\ref{1_2_contr}) through the rate 
$w ( k _ 1,k _ 2)$  of the metastable vacuum destruction by collisions of two particles. Namely, the relation is as follows
\be
\f {\G_ {2 \c}} {L} = L \, \int \f {d k _ 1 d k _ 2} {\l ( 2 \pi \r ) ^ 2} \, w (k _ 1, k _ 2) \, n(\vec k _ 1) n (\vec k _ 2) \, .
\label{2_p} 
\ee
The two-particle rate function $w$ is related to the dimensionless and Lorentz invariant $1+1$ dimensional analog $\s$ of a cross section in $3+1$-dimensions:  
\be
w (k _ 1, k _ 2)= {\s (k _ 1, k _ 2)} \, \times \,{{\rm {Flux}}}= {\s} \, \f {v _ {rel}} {L} \, ,
\ee  
where $v _ {rel}$ is the relative velocity of the two particles, which is expressed in terms of the energies of the particles and of the standard kinematical invariant
$I = \sqrt { \l ( k_1 \cdot k _ 2 \r ) ^2 - m^4 }$
as 
\be
v _ {rel} = \f {I} {E _ {k _ 1} E _ {k _ 2}} \, .
\ee

Due to the Lorentz invariance, the function $\s (k _ 1, k _ 2)$ can depend only on the invariant $I$, so that one can consider it in a form of a power series in $I$:
\be
\s(I) = \sum _ n c _ n I ^ n.
\label{sser}
\ee

In order to write the integral over the momenta of the two particles with proper integration limits one should take into account that the particles are identical and also that for the collision to take place they should move toward each other. Let particle `1' be the one on the left and the particle `2' on the right. Also let the spatial momentum $k$ be considered as positive when the particle moves from left to right. Then the collision takes place if and only if $k_1 > k_2 $, where both the spatial momenta can be of either sign, and the integral in Eq.(\ref{2_p}) takes the form
\be
\f {\G_ {2 \c}} {L} = \int_{-\infty}^{+\infty} {d k_1 \over 2 \pi \, E _ {k _ 1}} \, n(k_1) \, \int_{-\infty}^{k_1} {d k_2 \over 2 \pi \, E _ {k _ 2} } \, n(k_2) \, \s(I) \, I~.
\label{col_rate}
\ee
This integral can reproduce the two-particle thermal term in Eq.(\ref{1_2_contr}) only if the product $\s(I) I$ is constant, i.e. only if the expansion in Eq.(\ref{sser}) reduces to a single term $\s(I) = c_{-1}/I$. The integrand in Eq.(\ref{col_rate}) then factorizes into terms each depending on the absolute value of the momentum, so that the limits of integration can be rearranged as
\bea
\f {\G_ {2 \c}} {L} & = & c_{-1} \, \int_{-\infty}^{+\infty} {d k_1 \over 2 \pi \, E _ {k _ 1}} \, n(k_1) \, \int_{-\infty}^{k_1} {d k_2 \over 2 \pi \, E _ {k _ 2} } \, n(k_2) 
\nonumber \\
& = & \int_{-\infty}^{+\infty} {d k_1 \over 2 \pi \, E _ {k _ 1}} \, n(k_1) \, \int_0^{\infty} {d k_2 \over 2 \pi \, E _ {k _ 2} } \, n(k_2) = 2 c_{-1} \, \l [ \sum_{n=1}^{\infty} K_0(m \b n) \r ]^2~.
\eea
Upon a comparison with the expressions (\ref{D_act}) and (\ref{1_2_contr}) one thus arrives at the final result for the dimensionless invariant `cross section' for the destruction of the false vacuum in collision of two $\c$-particles:
\be
\s(I) = \f {\pi \eps} {2I} \,  \l [ 2 \pi g ^ 2 R ^ 2 I _ 0 ^ 2 (mR) \r ] ^ 2 \, \exp \l [ - \f {\pi \m ^ 2} {\eps} \r ]\, .
\label{t_prob}
\ee

\section{Direct semiclassical calculation for one particle}
In this section we consider another approach to calculating the rate of the false vacuum decay induced by one particle, which is similar to the one used in Ref.~\cite{mv2010} for the photon-induced Schwinger process. Interpreting the particle-induced decay rate as the width of the particle $\c$, one can  find it from the propagator of the $\c$ field in the background of the bounce as the imaginary part of the density-density correlator\footnote{This  is in a one to one correspondence with QED, where the attenuation rate of the photon is given by the imaginary part of current-current correlator.}:
\be
\Pi (t,x) = \langle \rho (t,x)\rho(0,0)\rangle.
\ee 
Using the formulas (\ref{e_act}) and (\ref{path_int}) we can rewrite the leading semiclassical term of the correlator in the following form
\bea
\mathrm{Im} \Pi (t,x) & = & \mathrm{Im} \int \mathcal {D} \vp \rho (t,x) \rho (0,0) \, e ^ {-S [\vp]} 
= \mathrm{Im} \int \mathcal{D}r \rho ^ {(r)} (t,x) \rho ^ {(r)} (0,0) e ^ {-S [r]} \nonumber \\
& = & \f {\G} {2 L} \int \rho ^ B (t-t_B,x-x_B) \rho ^ B (-t_B,-x_B) d ^ 2 x _ B,
\label{rho_corr}
\eea
with $\rho ^ B (t,x)$ from Eq.(\ref{b_dens}). 

The integration in Eq(\ref{rho_corr}) is performed in the following manner. We choose the coordinate system such that the coordinate $x$ in the integral is equal to zero. One can always do so by an appropriate rotation of the coordinates. Due to the  O$(2)$ symmetry, the full final result can be restored in an arbitrary coordinate system by simply promoting $t$ to $\sqrt{t ^ 2 + x ^ 2}$ in the final expression for the correlator. 
\begin{figure}[ht]
  \begin{center}
    \leavevmode
    \epsfxsize=7cm
    \epsfbox{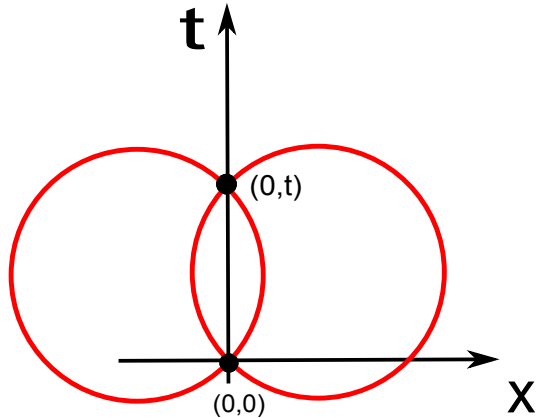}
    \caption{The two positions of the bounce for which the density correlator is not equal to zero.}
  \end{center}
\end{figure}
After this choice of orientation we make the change of the variables of integration 
\bea
z _ 1 & = & \sqrt{t _ B ^ 2 + x _ B ^ 2 }, \nonumber \\
z _ 2 & = & \sqrt{(t _ B - t ) ^ 2 + x _ B ^ 2 },
\label{2d_coord_c}
\eea
and rewrite the integral in the following form
\bea
g ^ 2 \int d t _ B d x _ B \delta \l ( \sqrt{t _ B ^ 2 + x _ B ^ 2 } - R \r ) 
\delta \l ( \sqrt{(t _ B - t ) ^ 2 + x _ B ^ 2 } - R \r )  \nonumber \\
= 2 g ^ 2 \int d z _ 1 d z _ 2 \delta(z _ 1 - R) \delta (z _ 2 - R) J\l ( \f {t _ B x _ B} {z _ 1 z _ 2}\r )
d z _ 1 d z _ 2 = \f {4 R ^ 2 g ^ 2} {t \sqrt{4 R ^ 2 - t ^ 2}},
\label{corrx}
\eea
where $J\l ( \f {t _ B x _ B} {z _ 1 z _ 2}\r )$ is the Jacobian of the transformation (\ref{2d_coord_c}). The factor of two in the second expression in Eq.(\ref{corrx}) comes from two possible configurations shown in Fig.3. After promoting $t \to \sqrt {t ^ 2 + x ^ 2} $ and making the Fourier transformation one finds
\be
\int d q _ 0 d q _ 1 \, \f {4 R ^ 2 g ^ 2} {\sqrt {t ^ 2 + x ^ 2} \sqrt{4 R ^ 2 - t ^ 2 - x ^ 2}} e ^ {i q _ 0 t + i q _ 1 x} = 
- 4 \pi ^ 2 g ^ 2 R ^ 2 I _ 0 ^ 2 (m R),
\ee
where the two-momentum $q$ is set on the mass shell, $q ^ 2 = - m ^ 2$. Using the formula (\ref{rho_corr}) 
one thus finds:
\be
\g _ \c = - \f {\mathrm{Im} \Pi } {m} = \f {\pi \, \eps} {m} \, g ^ 2 R ^ 2 I _ 0 ^ 2 (m R) 
\, \exp \l [ - \f {\pi \m ^ 2} {\eps} \r ],
\label{s_width}
\ee
which  coincides with Eq.(\ref{t_width}).

\section{Discussion and summary}

The formulas for the false vacuum decay rate induced by one (Eq.(\ref{t_width}) or Eq.(\ref{s_width})) and by a collision of two (Eq.(\ref{t_prob})) light particles of the weakly interacting field $\c$ merit some remarks. Firstly, one can readily notice that the width of the particle associated with the destruction of the vacuum is singular in the limit $m \to 0$, which is a consequence of the infrared behavior in this limit in 1+1 dimensions.  The singular growth at low $m$ however does not result in any unphysical artifacts, since the rate of decay for a state with any fixed finite energy $E$ is finite in the massless limit due to the time dilation factor $m/E$.
Secondly, at large $m R$ the exponential asymptotic behavior of the Bessel function correctly reproduces the known~\cite{sv,vs} semiclassical exponential factor in both the one-particle and two-particle processes: 
\be
\gamma _ \c \sim \exp \l [ - \f {\pi \m ^ 2} {\eps} + 2mR\r ]\, ,~~~~~~~w  \sim  \exp \l [ - \f {\pi \m ^ 2} {\eps} + 4 m R \r ]~.
\ee
It is worth noting that the enhancement factor in the collision is determined by the mass of the particles rather than by their c.m. energy $\sqrt{s}$. This behavior, which is in agreement with the geometrical calculation of the exponent in Refs.~\cite{sv,vs} is a consequence of the assumed linear in $\c$ interaction in Eq.(\ref{int_pot}). Had we assumed instead a quadratic in $\c$ interaction, the exponential factor would be determined by $\sqrt{s} \, R$ at $\sqrt{s} \gg m$, as is the case for the similar factor in the rate of destruction of metastable strings and walls by Goldstone bosons~\cite{mvsw}. Moreover, the absence of the dependence on the energy of the exponent in $\s$ is true only as long as the energy is small in comparison with the mass scale $M$ of the master field $\vp$, $\sqrt{s} \ll M$. The reason for this behavior is that iteration of the interaction in Eq.(\ref{int_pot}) through the short-distance propagator of $\vp$ gives rise to terms with higher powers of $\c$, which however are suppressed by inverse powers of $M$.

In summary. We have considered an illustrative 1+1 dimensional model of interacting scalar fields and demonstrated an application of the semiclassical thermal method to calculation of the rate of destruction of the false vacuum by one- and two- particle states of a light scalar field. For the one-particle state a direct calculation of the rate is also done in terms of the imaginary part of the particle propagator in the tunneling background. In the latter case the direct method is possibly simpler, while in the case of two particles the thermal approach appears to be more straightforward. Furthermore, the formula for the thermal decay rate produces a generating function for the probability of the process induced by an arbitrary number of particles.

The calculations discussed in the present paper can be readily generalized to other models of interaction and to models in higher dimensions.

\section*{Acknowledgments}
The work of A.M. is supported in part by the Stanwood Johnston grant from the Graduate
School of the University of Minnesota, RFBR Grant No. 07-02-00878 and by the Scientific
School Grant No. NSh-3036.2008.2. The work of M.B.V. is supported in part by the DOE
grant DE-FG02-94ER40823.


\end{document}